\documentclass[12pt]{iopart}
\usepackage{iopams}
\usepackage{graphicx}
\usepackage{epsfig}
\usepackage{color}
\usepackage{bm}

\begin{document}

\title[]{Physics of collisionless shocks - theory and simulation}

\author{A. Stockem Novo}
\address{Institut f\"ur Theoretische Physik, Lehrstuhl IV: Weltraum- \& Astrophysik, Ruhr-Universit\"at, Bochum, Germany}
\ead{anne@tp4.rub.de}

\author{A. Bret}
\address{ETSI Industriales, Universidad de Castilla-La Mancha, 13071 Ciudad Real, Spain\\
Instituto de Investigaciones Energ\'eticas y Aplicaciones Industriales, Campus Universitario de Ciudad Real, 13071 Ciudad Real, Spain}

\author{R. A. Fonseca\(^*\), L. O. Silva}
\address{GoLP/Instituto de Plasmas e Fus\~ao Nuclear, Instituto Superior T\'ecnico, Universidade de Lisboa, Lisboa, Portugal\\
\(^*\){\it also at:} DCTI, ISCTE - Lisbon University Institute Portugal}

\begin{abstract}
Collisionless shocks occur in various fields of physics. In the context of space and astrophysics they have been investigated for many decades. However, a thorough understanding of shock formation and particle acceleration is still missing. Collisionless shocks can be distinguished into electromagnetic and electrostatic shocks. Electromagnetic shocks are of importance mainly in astrophysical environments and they are mediated by the Weibel or filamentation instability. In such shocks, charged particles gain energy by diffusive shock acceleration. Electrostatic shocks are characterized by a strong electrostatic field, which leads to electron trapping. Ions are accelerated by reflection from the electrostatic potential. Shock formation and particle acceleration will be discussed in theory and simulations.
\end{abstract}

%\maketitle

\section{Introduction}

Collisionless shocks show interesting features for particle acceleration, which is why they are of importance in many fields of physics \cite{MedvedevApJ2005}-\cite{HF15}. They are generated by the interaction of charged particles with the surrounding self-generated fields -- in contrast to collision-dominated shocks, where two-particle interactions determine the physical behaviour. In a simple setup of two interpenetrating plasma slabs, either a strong electrostatic field \cite{ZR67} is generated due to the two-stream instability \cite{BD05}, or an electromagnetic field due to Weibel-like instabilities \cite{F59,W59}. The feedback of such fields mediates the shock formation process.

The Weibel or filamentation instability generate a turbulent field in which the charged particles are deflected. Parallel momentum is transferred into perpendicular momentum, and thus the particles are accumulated and the plasma slabs are compressed. In relativistic initially unmagnetised plasmas, the density compression is three times the initial density \cite{BM76}. The time scales of the shock formation are determined by the time scales of the electromagnetic instability \cite{SB15}. The details are given in Sec.\ \ref{sec:em}.

Electrostatic (ES) shocks form in plasmas with different components of electrons and (heavy) ions. A high mass and temperature difference is favourable for the generation of these shocks. An electrostatic potential is built-up which traps electrons in the downstream region and the electron distribution function is widened in parallel direction. The perpendicular components are less affected. This temperature anisotropy gives rise to the electromagnetic Weibel instability. The time scales of the formation of such electromagnetic modes are calculated in Sec.\ \ref{sec:es} and compared against the time scales of shock formation in order to determine their relevance during the shock formation process. We perform the analysis for plasmas with relativistic temperatures and fluid velocities. Laser-generated plasmas in the laboratory are now entering the relativistic regime, which is why it is important to explore the physics also in this parameter range.

\section{Initial setup for shock formation}

\begin{figure}[ht!]
\begin{center}
\includegraphics[width=10cm]{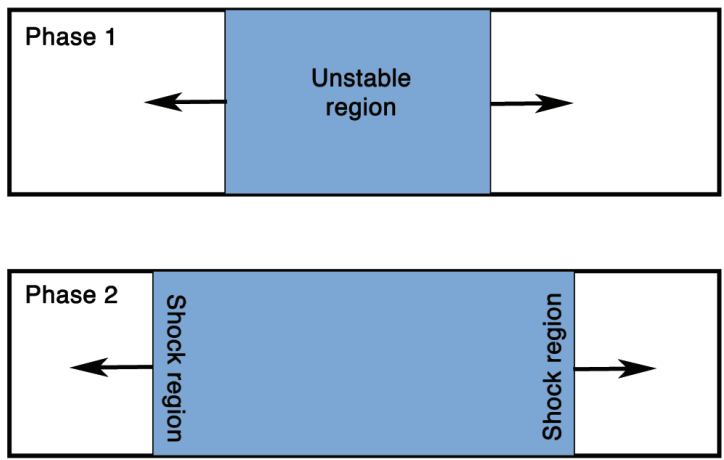}
\caption{Shock formation from the interaction of two counterstreaming beams. In Phase 1 the overlapping region becomes unstable. In Phase 2 two shocks are propagating into the upstream regions.}\label{fig:setup}
\end{center}
\end{figure}

We study shock formation in a simple symmetric setup (see Fig.\ \ref{fig:setup}). Let us consider two charge-neutral counterstreaming beams of electrons and positrons or ions with bulk velocities \(\pm v_0 \), temperatures \(T_e\), \(T_i\) and thermal parameter \(\mu = m c^2/k_B T\). The overlapping region turns unstable due to collisionless plasma instabilities (Phase 1), which will mediate two shocks propagating into the upstream regions (Phase 2).

%%%%%%%%%%%%%%%%%%%%%%%%%%%%%%%%%%%%%
%%%%%%%%%%%%%%%%%%%%%%%%%%%%%%%%%%%%%%%

The collisionless shock is of electromagnetic or electrostatic nature, depending on the initial parameters for the velocities and temperatures. We will discuss these details in the following sections.

\section{Shock formation in electromagnetic shocks}\label{sec:em}

Electromagnetic shocks develop if electromagnetic instabilities dominate in the overlapping region. In such a symmetric setup, the electromagnetic filamentation (Weibel) instability is the fastest mode for cold beams \(T_e \approx T_i \approx 0\) and relativistic velocities \(v_0 /c \approx 1\). The growth rate of the cold filamentation instability is given by \(\delta_a = \sqrt{2/ \gamma_0}\beta_0\, \omega_{pa} \) with \(\beta_0 = v_0/c\), Lorentz factor \(\gamma_0 = (1-\beta_0^2)^{-1/2}\) and plasma frequency of species \(a\) being \(\omega_{pa} = 4 \pi n_0 q_a^2/m_a\). The saturation time of this instability
\begin{equation}
\tau_{s,a} = \frac{1}{2 \delta_a} \ln \left( \frac{B_f^2}{B_i^2} \right)
\end{equation}
is a function of the initial and final magnetic field strengths. The initial magnetic field strength \(B_i\) is obtained from the evaluation of the spectra of spontaneous magnetic fluctuations, while the final field \(B_f \simeq \sqrt{8 \pi \gamma_0 n_0 mc^2}\) results from a trapping condition in the magnetic field structure \cite{BS13}. The saturation time in electron-positron shocks can then be expressed as
\begin{equation}
\tau_{s,e} \omega_{pe} = \frac{\sqrt{\gamma_0}}{2 \sqrt{2}} \ln \left( \frac{4}{15} \sqrt{\frac{6}{\pi}} n_0 \left(\frac{c}{\omega_{pe}}\right)^3 \mu \sqrt{\gamma_0} \right).
\end{equation}
For electron-positron pair shocks, the shock formation time is simply twice the saturation time of the instability in two dimensions, \(\tau_{f,e} = 2 \tau_{s,e}\), and a factor 3 in three dimensions, \(\tau_{f,e} = 3 \tau_{s,e}\) \cite{BS14}.

For electron-ion shocks the theory has to be extended by an extra term for the merging time of the filaments. At the saturation time of the filamentation modes in electron-positron plasmas, the transverse size of the filaments is already large enough in order to deflect particles strongly enough for efficient accumulation of particles. On the other hand, in electron-ion plasmas, the filaments are still on the electron scale. An additional merging time
\begin{equation}
 \tau_m \, \omega_{pi}= \frac{2^{3/2}}{\ln 2 }  \gamma_0^{1/2} \ln (m_i/m_e)
\end{equation}
is required in order to bring the filaments to the required size to significantly deflect the ions \cite{SB15}. The total shock formation time in electron-ion shocks is thus given by
\begin{equation}
\tau_{f,i} \, \omega_{pi} = (\tau_{s,i} + \tau_{m}) \, \omega_{pi} = 4.43 \,d\, \gamma_0^{1/2} \ln (m_i/m_e)
\end{equation}
with \(\tau_{s,i}\) the saturation time of the cold ion filamentation instability and \(d\) the number of dimensions. The predicted scaling is broadly consistent with particle-in-cell simulation results.

\section{Shock formation in electrostatic shocks}\label{sec:es}

\begin{figure}[ht!]
\begin{center}
\includegraphics[width=7cm]{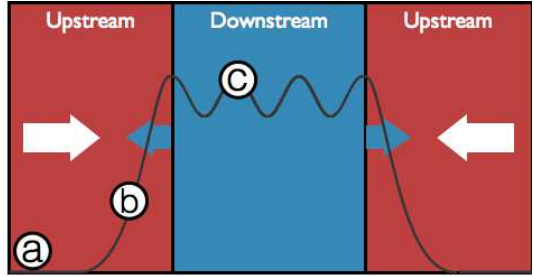}
\caption{An electrostatic shock forms in the interaction region of two counter propagating plasma slabs. The electrostatic potential \(\varphi\) increases monotonously from 0 in the upstream (a), a monotonous increase in the transition region (b) to its maximum \(\varphi_{max}\) in the downstream (c).}\label{fig2}
\end{center}
\end{figure}

Electrostatic shocks require a mass and temperature difference between ions and electrons. The electrostatic two-stream instability is dominant in the overlapping region, which is fast for low streaming velocities. A strong electrostatic potential is generated in the downstream region which can trap electrons (see Fig.\ \ref{fig2}), which can be approximated as a flat-top distribution. In the non-relativistic case, we follow Ref.Ê\cite{S72} for a steady-state solution of the shock and describe the electron distribution by a population of free streaming and a population of trapped electrons. If we introduce normalisations for the velocity, \(\beta = v/c\), and the electrostatic potential, \(\varphi = e \phi / m_e c^2\), electrons are free streaming if the condition \(|\beta_x| > \sqrt{\varphi}\) is fulfilled and trapped otherwise. The non-relativistic distribution of the electrons is then given by \cite{S72}
\begin{equation}\label{eq:nr}
	f_e = C_0  \left(\frac{\mu }{ 2 \pi} \right)^{3/2} 	e^{-\mu (\beta_y^2+\beta_z^2)/2}%
	\cases{\exp\{-\mu(\sqrt{\beta_x^2- 2 \varphi} + \beta_0)^2/2\} & $\quad \beta_x <-  \sqrt{2 \varphi}$\\
	\exp\{-\mu \beta_0^2/2\} & $\quad |\beta_x| \leq  \sqrt{2 \varphi}$\\
	 \exp\{-\mu(\sqrt{\beta_x^2- 2 \varphi} - \beta_0)^2/2\} & $\quad \beta_x >  \sqrt{2 \varphi}$\\}
\end{equation}
with the normalisation constant \( C_0 = \left[e^{\mu \varphi} \textrm{erfc} \sqrt{\mu \varphi} + 2  \sqrt{\mu \varphi/\pi} e^{-\mu \beta_0^2 / 2}\right]^{-1}\). Fig.\ \ref{fig2} demonstrates the ES shock configuration with an oscillatory potential in the shock downstream region.

%%%%%%%%%%%%%%%%%%%%%%%%%%%%%%%%%%%%%
%%%%%%%%%%%%%%%%%%%%%%%%%%%%%%%%%%%%%%%

\subsection{Relativistic generalisation of the distribution function}

For the relativistic generalisation of the electron distribution eq.\ (\ref{eq:nr}), we start with a Maxwell-J\"uttner distribution in the mean rest frame (subscript \(R\))
\begin{equation}
f_{re}^{R} = C_{R} \exp \left( - \mu_R \gamma_{R} \right),
\end{equation}
with the Lorentz factor always defined as \(\gamma = (1-\beta^2)^{-1/2}\), and perform a Lorentz transformation with \(\gamma_0\) into the moving (= laboratory) frame (subscript \(L\)) in \(\pm x\) directions. Thus, with \mbox{\(\gamma_R = \gamma_0 \gamma_L (1 \pm \beta_0 \beta_{L,x})\) } we obtain the relativistic distribution in the laboratory frame
\begin{equation}\label{eq:labdist}
f_{re}^{L} = C_L  \exp \left( -\mu_L \gamma_L \left( 1\pm \beta_0 \beta_{L,x} \right) \right)
\end{equation}
with \(\mu_L = \mu_R \gamma_0\). \(C_R\) and \(C_L\) are normalisation constants in the mean rest frame and in the laboratory frame, respectively.

We now introduce the dependency on the electrostatic potential \(\varphi\). From energy conservation, we obtain a balance between the kinetic energy in the upstream, where the bulk is moving with \(\gamma_L\), and the total energy in the shock transition region, where kinetic energy is transformed into potential energy. The bulk Lorentz factor is reduced to \(\gamma'_L\). This is expressed by
\begin{equation}\label{eq:energycons}
m_e c^2 (\gamma_L-1) = m_e c^2 (\gamma_L'-1 ) - e \phi 
\end{equation}
or simply \(\gamma_L = \gamma_L' - \varphi\). Electrostatic shocks are mainly one-dimensional, which is why we assume that the longitudinal and transverse processes can be separated. Introducing the dimensionless momentum \(u = \beta \gamma\), we approximate \(u_{L,x} \approx \sqrt{( \sqrt{1+u_{L,x}'^2} -\varphi)^2-1}\). We can then write the relativistic generalisation of  the electron distribution in eq.\ (\ref{eq:nr}), if we replace \(\gamma_L\) and \(\beta_{L,x}\) in eq.\ (\ref{eq:labdist}) by their \(\varphi\) dependent terms and with \(\gamma:= \gamma_L' \) and \(u_x := u_{L,x}' \) we obtain
\begin{equation}\label{eq:reldist}
	f_{re} (\mathbf u)= C_{r0} %
	\cases{\exp \left\{-\mu \left[ \gamma_0 \left( \gamma - \varphi \right) -1 + u_0 \sqrt{(\sqrt{1+u_x^2}-\varphi)^2-1} \right] \right\}	 &$\quad u_x < - u_c$\\
	\exp \left\{-\mu \left[ \gamma_0 \gamma_\perp -1 \right] \right\} & $\quad |u_x| \leq u_c$\\
	\exp \left\{-\mu \left[ \gamma_0 \left( \gamma - \varphi \right) -1 - u_0 \sqrt{(\sqrt{1+u_x^2}-\varphi)^2-1} \right] \right\} & $\quad u_x > u_c$\\}
\end{equation}
with the definitions \(\gamma = \gamma(\mathbf u) = \sqrt{1+ \mathbf u^2 }\) and \(\gamma_\perp = \sqrt{1+ u_\perp^2 }\). 
The definition of the trapping velocity \(u_c\) is derived from the balance of electrostatic and kinetic energy in eq.\ (\ref{eq:energycons}) for \(\gamma_L = 1\) which gives \(\gamma=\gamma_c := 1+ \varphi\) and \(u_c = \sqrt{\gamma_c^2-1}\). The non-relativistic approximation of this trapping velocity agrees with the non-relativistic definition in eq.\ (\ref{eq:nr}) for \(u_c \rightarrow \beta_c =  \sqrt{2\varphi}\) for \(\beta_c \ll 1\).

The normalisation constant \(C_{r0}\) is obtained from
\begin{equation}
 \int d^3 u \, f_{re} (\mathbf u) = 1
\end{equation}
with
\begin{eqnarray}
	C_{r0} &=& \frac{\gamma_0^2\mu^2}{2 \pi e^\mu} \left[ 2u_c (1+\gamma_0 \mu) e^{-\mu \gamma_0} + e^{\mu \gamma_0 \varphi} \sum_{\pm} \int_{g_c}^\infty
	d \gamma \frac{\gamma (1+\mu \gamma_0 \gamma)}{\sqrt{\gamma^2-1}}  e^{[-\mu(\gamma_0 \gamma \pm u_0 \sqrt{(\gamma-\varphi)^2-1})]}\right]^{-1}.
\end{eqnarray}

\begin{figure}[ht!]
\begin{center}
\includegraphics[width=15cm,origin=center]{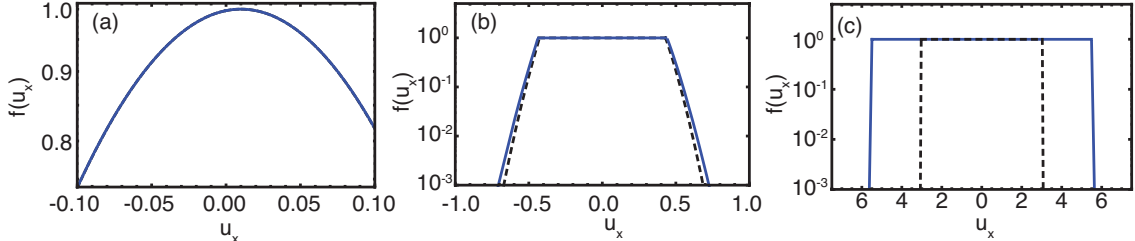}
\vspace{-18pt}
\caption{Electron distributions for \(\varphi = 0\) (a), \(\varphi_{refl} = 0.09\) (b) and \(\varphi_{max} = 4.58\) (c) for \(u_0 = 0.01\) and \(\mu = 50\) obtained from the relativistic expression eq. (\ref{eq:reldist}) given in blue and the non-relativistic approximation eq. (\ref{eq:nr}) in dashed black.}\label{fig3}
\end{center}
\end{figure}

Eqs.\ (\ref{eq:nr}) and (\ref{eq:reldist}) describe the electron distribution in the quasi-steady ES shock. In fig.\ \ref{fig3} we show the change of the distribution function from the initial Maxwell distribution for \(\varphi = 0\) to a flat-top distribution for \(\varphi > 0\) for \(u_0 = \beta_0 \gamma_0 = 0.01\) and \(\mu = 50\), corresponding to \(k_B T_e = 10 \) keV. The relativistic expression eq.\ (\ref{eq:reldist}) is also compared against the non-relativistic expression eq.\ (\ref{eq:nr}). In the far upstream, the distribution is Maxwellian, see fig.\ \ref{fig3}a.
In the shock transition region, where \(0 < \varphi \leq \varphi_{max}\), the distribution \(f(u_x)\) broadens and becomes flat-top. \(\varphi_{max}\) denotes the saturation value of the electrostatic potential in the far downstream which can be obtained by solving numerically the differential equation \(\frac{1}{2} \left(\frac{\partial \varphi}{\partial x} \right)^2 + \Psi(\varphi) = 0\) with the  Sagdeev potential \(\Psi(\varphi)\) \cite{TK71,SM06}. Fig.\ \ref{fig3}b shows the distribution function at the ion reflection condition, where the kinetic energy of the ions equals the electrostatic energy. In normalised quantities, this is given by \(\varphi=\varphi_{refl} := (\gamma_0 -1 ) m_i / m_e\). In the far downstream, for \(\varphi= \varphi_{max}\) the non-relativistic description breaks since \(\varphi_{max} > 1\).

%%%%%%%%%%%%%%%%%%%%%%%%%%%%%%%%%%%%%
%%%%%%%%%%%%%%%%%%%%%%%%%%%%%%%%%%%%%%

In this configuration, the electrons are usually treated kinetically, while the higher inertia ions are described with a fluid model. In the following section, we summarise the initial conditions for which an ES shock forms.

\subsection{Conditions for ES shock formation}

Here, we summarise the results on ES shock formation conditions. In Refs.\ \cite{FS70,FF71} a condition for the maximum Mach number was found as \(1 < M_{max} \lesssim 3.1\), with the Mach number \(M\) defined as the ratio of the upstream velocity in the shock rest frame to the ion sound speed, \(M = v_0'/c_s = (v_0 + v_{sh}) / c_s\). The shock velocity in the laboratory frame can be approximated from the shock jump conditions as \(v_{sh} /c = \sqrt{\gamma_{ad}-1} \sqrt{(\gamma_0-1)/(\gamma_0+1)}\) with the ideal gas adiabatic constant \(\gamma_{ad}\) \cite{BM76}. The ion sound speed is given by \(c_s = \sqrt{k_B T_e / m_i}\). This can be generalised for relativistic plasmas with the relativistic Mach number \(\mathcal{M} = u_0' / u_s\), where \(u_0' = \beta_0' \gamma_0'\) is the dimensionless momentum in the shock rest frame and \(u_s = \beta_s \gamma_s\) with \(\beta_s = \sqrt{k_B T_e / m_i c^2}\) \cite{FS70,SF13e}.
This imposes a condition for the upstream fluid velocity and electron temperature
\begin{equation}\label{eq:sfcond}
 1 < u_0' \sqrt{\frac{m_i}{m_e}} \sqrt{\mu} \leq 3.1.
\end{equation}

\section{Calculation of unstable modes}\label{sec:modes}

During the early stage of ES shock formation the broadening of the electron distribution, which we observed in the previous section, occurs mainly in the longitudinal direction. The transverse directions stay almost unaffected. This has also been observed by particle-in-cell simulations \cite{SG14}. The generated temperature anisotropy in the electron distribution gives rise to electromagnetic modes.

We develop now a model to describe the growth rate of the electromagnetic modes in such a setup. Later, we will compare them to the shock formation time scales and the growth rate of the cold ion-ion instability in order to determine their relevance.

\subsection{Dispersion relation of EM waves in relativistic plasmas}

We start from the electron distribution in eq.\ (\ref{eq:reldist}) in order to evaluate the dispersion relation of electromagnetic waves in plasma
\begin{equation}\label{disprel}
	k^2 c^2 - \omega^2 - \omega_{pe}^2 (U + V) = 0,
\end{equation}
with
\begin{equation}
	U = \int_{-\infty}^\infty d^3u \frac{u_x}{\gamma} \frac{\partial f}{\partial u_x}
\end{equation}
and
\begin{equation}
	V = \int_{-\infty}^\infty d^3u \frac{u_x^2}{\gamma \left( \gamma \frac{\omega}{kc}- u_z \right)} \frac{\partial f}{\partial u_z}.
\end{equation}
We consider only fluctuations \(\mathbf k = k \mathbf{e_z}\) perpendicular to the fluid velocity \(\mathbf{u_0} = u_0 \mathbf{e_x}\) in order to simplify the geometry. An evaluation of the integrals leads to
\begin{eqnarray}\label{eq:relu}
	U &=& - C_{r0} 2 \pi \mu e^{\mu (\gamma_0 \varphi +1 )} \sum_\pm \int_{\gamma_c}^\infty \sqrt{\gamma^2 -1} e^{\mp \mu u_0 \sqrt{(\gamma - \varphi)^2-1}} \nonumber \\
	&& \times \left[ \gamma_0 \gamma \Gamma \left( 0,\mu \gamma_0  \gamma \right) \pm \frac{\beta_0}{\mu} \frac{\gamma - \varphi}{\sqrt{(\gamma - \varphi)^2-1}} e^{-\mu \gamma_0 \gamma} \right]
\end{eqnarray}
and
\begin{eqnarray}\label{eq:relv}
	V &=& C_{r0} 4 \mu \gamma_0 \int_0^\infty du_z u_z^2 \int_0^\infty du_y  \left[  \frac{2}{\gamma_\perp} e^{-\mu (\gamma_0 \gamma_\perp -1)}   \int_0^{u_c} d u_x \frac{u_x^2}{\gamma (\gamma^2 y^2 + u_z^2)}  \right. \nonumber \\
	&& + \left.  \sum_\pm e^{\mu (\gamma_0 \varphi + 1)} \int_{u_c}^\infty \frac{u_x^2}{\gamma^2 (\gamma^2 y^2 + u_z^2)} e^{-\mu( \gamma_0 \gamma \pm u_0 \sqrt{(\sqrt{1+u_x^2}-\varphi)^2-1}) } \right]
\end{eqnarray}
where the last integrations have to be done numerically for the general case.

\begin{figure}[ht!]
\begin{center}
\includegraphics[width=13cm]{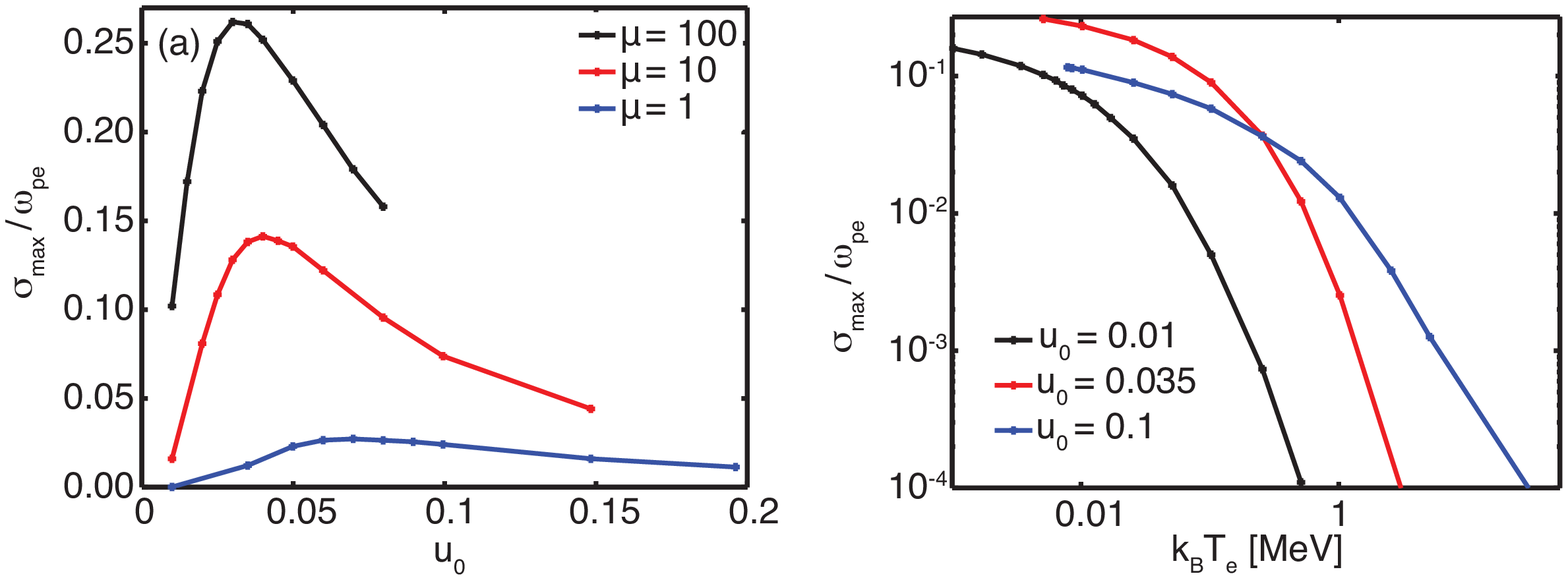}
\vspace{-18pt}
\caption{Maximum growth rate \(\sigma_{max}\) vs.\ \(u_0\) (a) and \(k_B T_e\) (b) obtained from dispersion relation eq.\ (\ref{disprel}) with relativistic expressions eqs.\ (\ref{eq:relu}) and (\ref{eq:relv}).}\label{fig4}
\end{center}
\end{figure}

We use eqs.\ (\ref{eq:relu}) and (\ref{eq:relv}) to solve the dispersion relation (\ref{disprel}) numerically to obtain the growth rate \(\sigma(k) := \Im (\omega (k))\). In Figs.\ \ref{fig4}a and b we plot the maximum growth rate \(\sigma_{max} := \max(\sigma(k))\) for different values of \(\mu\) and \(u_0\). The maximum growth rate was evaluated at the ion reflection condition \(\varphi_{refl}= (\gamma_0-1) m_i/m_e\). The role of the potential will be discussed in sec.\ \ref{sec:potential}.

%%%%%%%%%%%%%%%%%%%%%%%%%%%%%%%%%%%%%
%%%%%%%%%%%%%%%%%%%%%%%%%%%%%%%%%%%%%%%
%

\(\sigma_{max}\) shows a maximum as a function of the velocity. The higher the electron temperature, i.e. the lower \(\mu \propto T_e^{-1}\), the higher the velocity at which the maximum of \(\sigma_{max}\) appears (fig.\ \ref{fig4}a). On the other hand, \(\sigma_{max}\) increases with \(\mu\). For an easier interpretation, we plotted the dependence of \(\sigma_{max}\) against the electron temperature in Fig.\ \ref{fig4}b. The higher the initial electron temperature, the lower is the growth rate \(\sigma_{max}\).

\subsection{Dispersion relation of EM waves in non-relativistic plasmas}\label{sec:nonrel}

\begin{figure}[ht!]
\begin{center}
\includegraphics[width=8cm,origin=c]{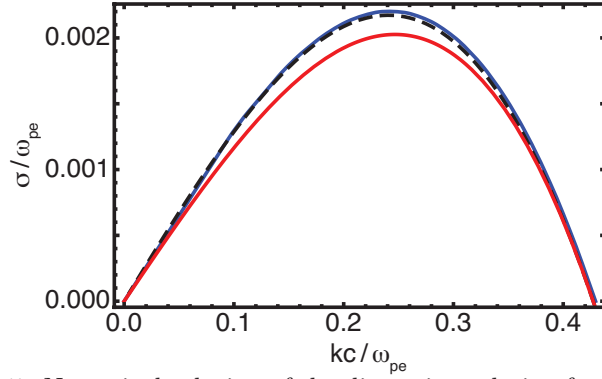}
\vspace{-18pt}
\caption{Numerical solution of the dispersion relation for the relativistically correct eq.\ (\ref{eq:nrelv}) in blue, the non-relativistic approximations from eq.\ (\ref{disprel:nr}) in black and from eq.\ (\ref{eq:sigma}) in red for \(\beta_0 = 0.01\), \(\mu = 100\) and \(\varphi = 0.05 \,(\gamma_0-1) \, m_i/m_e\).}\label{fig5}
\end{center}
\end{figure}

In order to further discuss the properties of this model and to check the consistency with previous models, dispersion relation (\ref{disprel}) is approximated for non-relativistic velocities, i.e. \(\beta_0 \ll 1\) and \(\mu \gg 1\). We obtain \(U_n \approx -1\) and
\begin{equation}\label{eq:nrelv}
 V_n \approx C_0 \left[  1+ \frac{\omega}{kc} \sqrt{\frac{\mu}{2}} \textrm Z \left(  \frac{\omega}{kc} \sqrt{\frac{\mu}{2}}  \right) \right] \left \{ e^{\mu \varphi} \textrm{erfc} \sqrt{\mu \varphi} + 2 \sqrt{\frac{ \mu\varphi}{\pi}} + \frac{4}{3} \sqrt{ \frac{\mu^3 \varphi^3}{\pi} } e^{-\mu \beta_0^2 / 2}  \right\}.
\end{equation}
 For zero beam velocity and \(\varphi = 0\), this is exactly the well-known solution of the Weibel instability \cite{W59}. For small frequencies \(\omega  \ll kc\), the plasma dispersion function \(Z\) can be further approximated and we get
\begin{equation}
 V_e \approx C_0 \left[  1+ \frac{\imath \omega}{kc} \sqrt{\frac{\mu \pi}{2}} -\mu \frac{\omega^2}{k^2 c^2} \right] \left \{ e^{\mu \varphi} \textrm{erfc} \sqrt{\mu \varphi} + 2  \sqrt{\frac{\mu\varphi}{\pi}} + \frac{4}{3} \sqrt{ \frac{\mu^3 \varphi^3}{\pi} } e^{-\mu \beta_0^2 / 2}  \right\}.
\end{equation}
The dispersion relation then reads 
\begin{equation}\label{disprel:nr}
	k^2c^2-\omega^2+ \omega_{pe}^2 \left[ 1- V(\varphi ) \left( 1 + \imath \frac{\omega}{kc} \sqrt{\frac{\pi \mu}{2}} \right)  \right] =0
\end{equation}
with \(V(\varphi )= C_0  \left \{ e^{\mu \varphi} \textrm{erfc} \sqrt{\mu \varphi} + 2 \sqrt{\frac{ \mu \varphi}{\pi}} + \frac{4}{3} \sqrt{ \frac{\mu^3 \varphi^3}{\pi} } e^{-\mu \beta_0^2 / 2}  \right\}\). The solution of the dispersion relation can now be derived analytically, which is given by
\begin{equation}\label{eq:sigma}
	\sigma (k) = \Im(\omega(k)) \approx \sqrt{\frac{2}{\mu \pi}} k c \left[ 1- \frac{k^2 c^2 + \omega_{pe}^2}{\omega_{pe}^2V(\varphi )} \right]
\end{equation}
with \(k_0^2c^2 = \omega_{pe}^2(V(\varphi)-1)/3\) the location of the maximum and the maximum growth rate 
\begin{equation}\label{eq:sigmamax}
	\sigma_{max} \approx \sqrt{\frac{1}{\pi \mu}}\frac{\omega_{pe}}{V(\varphi)} \left( \frac{2}{3} (V(\varphi )-1)  \right)^{3/2}.
\end{equation}
For comparison, fig.\ \ref{fig5} shows the relativistically correct solution of the dispersion relation, obtained from eqs.\ (\ref{eq:relu}) and (\ref{eq:relv}), and the non-relativistic approximations from eqs.\ (\ref{disprel:nr}) and (\ref{eq:sigma}). The parameters used are \(\beta_0 = 0.01\), \(\mu = 100\) and \(\varphi = 0.05 \,(\gamma_0-1) \, m_i/m_e\). The maximum growth rate matches with the approximation in eq.\ (\ref{eq:sigmamax}), \(\sigma_{max} = 2.0 \times 10^{-3} \omega_{pe}\) and the location of the maximum at \(k_0 = 0.25 \, \omega_{pe}/c\).

%%%%%%%%%%%%%%%%%%%%%%%%%%%%%%%%%%%%%
%%%%%%%%%%%%%%%%%%%%%%%%%%%%%%%%%%%%%%%
%

\subsection{The role of the electrostatic potential}\label{sec:potential}
\begin{figure}[ht!]
\begin{center}
\includegraphics[width=8cm,origin=c]{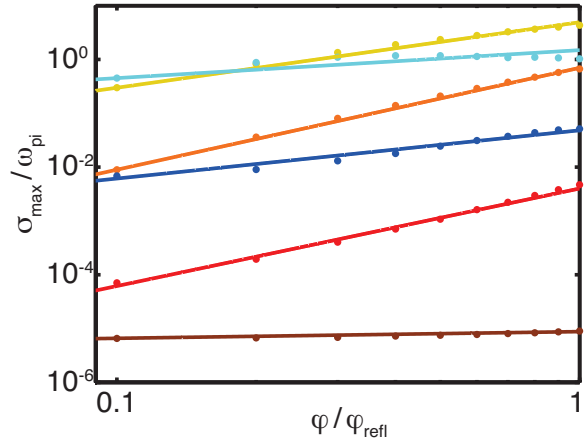}
\caption{Maximum growth rate for \(u_0 = 0.01\) and \(\mu = 100\) (yellow), 10 (orange), 1 (red), 0.1 (brown) and \(u_0 = 0.1\) and \(\mu = 1\) (light blue), 0.1 (dark blue) with \(\varphi \in [0, \varphi_{refl}]\).}\label{fig:pot}
\end{center}
\end{figure}

The electrostatic potential \(\varphi\) is quickly built up during the electrostatic shock formation. Nevertheless, the electromagnetic growth rate \(\sigma(k)\) depends on \(\varphi\) and thus, the dependence should be investigated. We calculate the growth rate numerically for values \( 0 \leq \varphi \leq \varphi_{refl}\), which are plotted in fig.\ \ref{fig:pot}.

%%%%%%%%%%%%%%%%%%%%%%%%%%%%%%%%%%%%%
%%%%%%%%%%%%%%%%%%%%%%%%%%%%%%%%%%%%%%%
%

We find a power-law dependence of the growth rate on the electrostatic potential, which gives
\begin{equation}
\sigma_{max} \propto \varphi^\alpha
\end{equation}
with \(\alpha > 0\). This \(\varphi\) dependence can be interpreted as the changing growth rate across the steady-state shock. If we assume, as a simple approach, that the potential grows linear in time, we can connect the growth rate \(\sigma_{max}\) with time, which will be further discussed in the next section.

\section{Comparison of different time scales}\label{sec:comparison}

In order to determine the relevance of the electromagnetic electron Weibel modes during the ES shock formation process, we compare the growth rate given by eq.\ (\ref{disprel}), which we label from now on \(\sigma_{EM,ee}\), with other processes.

\subsection{ES Shock formation time}

The time scales of ES shock formation are determined by the growth rate of the ion-ion electrostatic instability which is given by \cite{BD05}
\begin{equation}\label{eq:esii}
\sigma_{ES,ii} = \frac{1}{2\gamma_0^{3/2}}\omega_{pi}.
\end{equation}
%.
Estimating the shock formation time with 5 times the inverse maximum growth rate, we obtain \(t_{sf} = 10 \gamma_0^{-3/2} \, m_i/m_e \, \omega_{pe}^{-1}\), which was confirmed in simulations (see \cite{SF14}).

\subsection{Growth rate of the electromagnetic cold ion-ion instability}

Another competing process in ES shocks is the cold ion-ion filamentation instability, which has a growth rate \cite{CP98}
\begin{equation}\label{eq:emii}
\sigma_{EM,ii} = \beta_0\sqrt{\frac{2}{\gamma_0}} \omega_{pi}.
\end{equation}
The EM mode of the ion-ion instability (\ref{eq:emii}) grows faster than the ES mode  (\ref{eq:esii}) for fluid velocities \(v_0/c > 1/3\). In the parameter range \(v_0 / c \leq 0.1\), which we looked on in this paper, the ion-ion EM modes can be neglected.

\subsection{Dominant regimes}

A comparison of the ES shock formation time scales with the EM modes makes it possible to determine parameter regimes, for which a shock stays electrostatic. For this, we compare the growth rate from eq.\ (\ref{disprel}) with eq.\ (\ref{eq:esii}) for different electron temperatures, expressed by \(k_B T_e / m_e c^2\), and fluid velocities, given by \(u_0 = \beta_0 \gamma_0\). For simplicity, we choose an electrostatic potential \(\varphi = \varphi_{refl}\).

\begin{figure}[ht!]
\begin{center}
\includegraphics[width=9cm,origin=c]{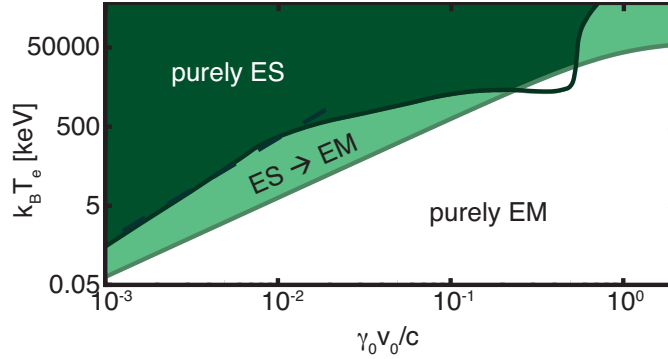}
\vspace{-18pt}
\caption{Determination of dominant regimes as a function of dimensionless momentum \(u = \beta_0 \gamma_0\) and electron temperature in keV, cf. Ref.\ \cite{SF14}.}\label{fig:regimes}
\end{center}
\end{figure}

In the dark green region in fig.\ \ref{fig:regimes} the growth rate of the EM electron instability \(\sigma_{EM,ee}\) is smaller than the ES ion-ion growth rate \(\sigma_{ES,ii}\). We label this domain the purely ES domain. In contrast, in the light green region, an ES shock will turn EM since \(\sigma_{ES,ii} < \sigma_{EM,ee}\). In the white region (EM), no electrostatic shock develops because condition (\ref{eq:sfcond}) is not fulfilled. This regime is electromagnetically dominated from the beginning, with shock formation according to section \ref{sec:em}.

%%%%%%%%%%%%%%%%%%%%%%%%%%%%%%%%%%%%%
%%%%%%%%%%%%%%%%%%%%%%%%%%%%%%%%%%%%%%%

\begin{figure}[ht!]
\begin{center}
\includegraphics[width=8cm,origin=c]{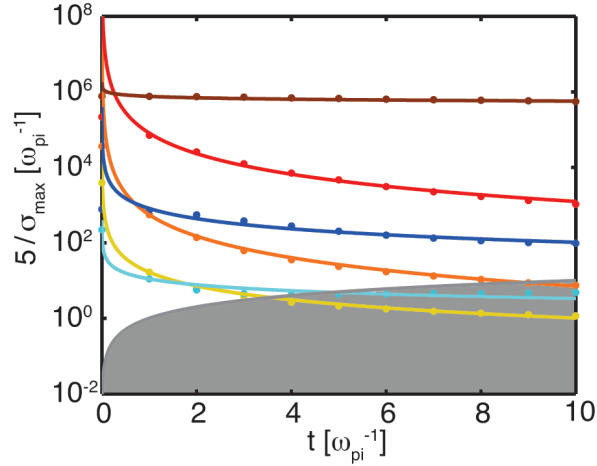}
\caption{Characteristic time \(t_{char} \approx 5/\sigma_{max}\) against time during shock formation \(t \in [0, 10 \gamma_0^{-3/2} \omega_{pi}^{-1}]\) with same colour coding as in fig.\ \ref{fig:pot}. The grey region shows \(t_{char} \leq t\), where EM modes are important during ES shock formation.}\label{fig:gr}
\end{center}
\end{figure}

A more detailed analysis of the scenario in fig.\ \ref{fig:regimes} has been done in fig.\ \ref{fig:gr} for different values of the electrostatic potential \(\varphi\). For different fluid velocities \(u_0\) and temperature parameters \(\mu = m_e c^2/k_B T_e\), the maximum growth rate \(\sigma_{max}\) is calculated as in fig.\ \ref{fig:pot}. The characteristic time is then calculated as \(t_{char} \approx 5/\sigma_{max}\) and plotted against the time of shock formation, where we assumed that the potential grows linearly as \(\varphi \propto t\).

%%%%%%%%%%%%%%%%%%%%%%%%%%%%%%%%%%%%%
%%%%%%%%%%%%%%%%%%%%%%%%%%%%%%%%%%%%%%%
%

The grey area in fig.\ \ref{fig:gr} shows the region, where EM modes develop faster than the ES shock, which is the case e.g. for \(u_0 = 0.01\) for \(\mu \gtrsim 10\) (\(\lesssim \) 51 keV) or for \(u_0 = 0.1\) for \(\mu \gtrsim 0.1\) (\(\lesssim \) 5 MeV). Fig.\ \ref{fig:gr} also shows that a detailed  analysis of the growth rate is not necessary, since the characteristic time \(t_{char}\) quickly saturates. The rough estimate in fig.\ \ref{fig:pot} is thus sufficient for a qualitative determination of the regimes.

\section{Summary and conclusions}

Electromagnetic and electrostatic shocks can both develop from the same simple symmetric setup of counterstreaming beams. The choice of the initial plasma parameters determines the final shock character.

Electromagnetic shocks develop in plasmas with large beam velocities and low temperatures where the electromagnetic filamentation instability is the fastest mode. In electron-positron pair plasmas, the shock formation time is simply twice the saturation time of the instability in 2D. In electron-ion plasmas an extra merging time is necessary in order to obtain the right scale of the filaments to efficiently scatter the ions. Thus, the shock formation in electron-ion plasmas is delayed by almost a factor 3 compared with the pair shock, \(\tau_{f,i} \omega_{pi} \approx 3 \tau_{f,e} \omega_{pe}\).

Electrostatic shocks form in electron-ion plasmas with small beam velocities and large electron temperatures. Here, the electrostatic two-stream instability dominates. An electrostatic potential is generated in which electrons are trapped. The subsequent deformation of the electron distribution gives rise to electromagnetic Weibel modes which can destroy the electrostatic shock features.

\section*{Acknowledgements}
This work was supported by the European Research Council (ERC-2010-AdG grant 267841), grant ENE2013-45661-C2-1-P from the Ministerio de Educaci\'on y Ciencia, Spain, and grant PEII-2014-008-P from the Junta de Comunidades de Castilla-La Mancha. The authors acknowledge the Gauss Centre for Supercomputing (GCS) for providing computing time through the John von Neumann Institute for Computing (NIC) on the GCS share of the supercomputer JUQUEEN at J\"ulich Supercomputing Centre (JSC).

\end{document}